%% file: main.tex
\documentclass[runningheads]{llncs}
\usepackage{graphicx}
\usepackage{amsmath}

\usepackage[labelfont=bf]{caption}                         % set caption font size
\usepackage[subrefformat=parens]{subfig}
\usepackage{enumitem}  
\usepackage{amssymb}
\usepackage[misc]{ifsym}
\usepackage{graphicx}
\usepackage{threeparttable}
\usepackage{ntheorem}
\usepackage{color}
\usepackage{colortbl}
\usepackage{multirow}
\usepackage{amsmath}
\usepackage{comment}
\usepackage{bm}                                            % bold math font
\usepackage{etoolbox}                                      % \newtoggle, \toggletrue, \iftoggle
\usepackage{url}
\usepackage{nth}
\usepackage{cite}
\usepackage{balance}
\usepackage[bookmarks=false]{hyperref}                     %
\usepackage{courier}                                       % courier font, used in \texttt 
\usepackage{listings}                                      % typeset source code listings
\usepackage[]{algpseudocode}                               % algorithm package
\algrenewcommand\textproc{\texttt}
\makeatletter\let\c@float@type\relax\makeatother
\makeatletter\let\float@addtolists\relax\makeatother
\usepackage{algorithm}
\hypersetup{
    colorlinks = true,
    citecolor  = blue,
    linkcolor  = blue,
    urlcolor   = blue,
}

\makeatletter
\newcommand{\thickhline}{%
	\noalign {\ifnum 0=`}\fi \hrule height 1pt
	\futurelet \reserved@a \@xhline
}

\graphicspath{{./figs/}}

\begin{document}

\title{MemWarp: Discontinuity-Preserving Cardiac Registration with Memorized Anatomical Filters}

\author{
Hang Zhang \inst{1} \Letter    \and % Zhang, Hang
Xiang Chen \inst{2} \and % Wang, Rongguang
Renjiu Hu \inst{1} \and % Hu, Renjiu
Dongdong Liu \inst{3} \and % Liu, Dongdong
Gaolei Li \inst{4}  \and % Li, Jiahao
Rongguang Wang \inst{5} % Wang, Rongguang
}
% \index{Zhang, Hang} 
% \index{Chen, Xiang}
% \index{Hu, Renjiu}
% \index{Liu, Dongdong}
% \index{Wang, Rongguang}
% \index{Li, Gaolei}

\authorrunning{H. Zhang \emph{et al}.}

\institute{Cornell University \and
Hunan University \and
New York University  \and 
Shanghai Jiao Tong University  \and 
University of Pennsylvania \\
\email{\textbf{hz459@cornell.edu}}
}

\maketitle

\input{./docs/abstract}
\input{./docs/introduction}
\input{./docs/method}
\input{./docs/experiment}

\input{./docs/conclusion}

\bibliographystyle{splncs04}
\bibliography{mybibliography}

\end{document}

%% file: docs/abstract.tex
\begin{abstract}

Many existing learning-based deformable image registration methods impose constraints on deformation fields to ensure they are globally smooth and continuous. 
However, this assumption does not hold in cardiac image registration, where different anatomical regions exhibit asymmetric motions during respiration and movements due to sliding organs within the chest.
Consequently, such global constraints fail to accommodate local discontinuities across organ boundaries, potentially resulting in erroneous and unrealistic displacement fields.
In this paper, we address this issue with \textit{MemWarp}, a learning framework that leverages a memory network to store prototypical information tailored to different anatomical regions. 
\textit{MemWarp} is different from earlier approaches in two main aspects: firstly, by decoupling feature extraction from similarity matching in moving and fixed images, it facilitates more effective utilization of feature maps; secondly, despite its capability to preserve discontinuities, it eliminates the need for segmentation masks during model inference.
In experiments on a publicly available cardiac dataset, our method achieves considerable improvements in registration accuracy and producing realistic deformations, outperforming state-of-the-art methods with a remarkable 7.1\% Dice score improvement over the runner-up semi-supervised method.
Source code will be available at \url{https://github.com/tinymilky/Mem-Warp}.

\keywords{Deformable image registration \and Memory network \and Discontinuity preserving}

\end{abstract}

%% file: docs/introduction.tex
\section{Introduction}

Cardiovascular disease, a major cause of death worldwide \cite{timmis2022european}, depends on medical imaging, especially cine-MRI \cite{khalil2018overview}, for diagnosis and treatment.
Deformable image registration \cite{balakrishnan2019voxelmorph}, a crucial step for cardiac analysis, has seen improvements through learning-based neural networks. 
These models vary from unsupervised to semi- and weakly-supervised frameworks. Unsupervised methods are favored for their simplicity, requiring only raw images for training and inference. 
In contrast, semi-supervised methods need segmentation masks during training, while weakly-supervised models require them in both training and inference phases.

Learning-based registration models \cite{balakrishnan2019voxelmorph,mok2020large,chen2022transmorph} outperform traditional iterative optimization approaches \cite{beg2005computing,ashburner2007fast,vercauteren2009diffeomorphic,avants2011reproducible,marstal2016simpleelastix} in efficiency and precision. 
Yet, they often presuppose globally smooth deformations, a premise that doesn't align with the dynamic nature of cardiac motions influenced by heartbeat and respiratory-induced organ sliding. 
Additionally, volume shifts between end-diastole (ED) and end-systole (ES) phases, such as expansion of the left ventricular myocardium (LVM) and reductions in the left ventricular blood pool (LVBP) and right ventricle (RV), underscore the need for models that can handle local discontinuities to accurately depict cardiac motions.

Despite the clear need for discontinuity-preserving methods to capture the cardiac cycle's complexity, the field remains underexplored. 
Ng et al.~\cite{ng2020unsupervised} pioneered this area by integrating a discontinuous regularizer for local discontinuity without segmentation masks in an unsupervised manner, though accurately defining organ boundaries remains challenging. 
DDIR \cite{chen2021deep} and textCSF \cite{chen2024spatially} address this by using segmentation masks to refine boundaries in a weakly-supervised manner, yet they require segmentation masks during both training and inference, making registration accuracy highly dependent on the quality of segmentation.

To tackle these challenges, we introduce MemWarp, a semi-supervised framework that balances local smoothness with the preservation of local discontinuities.
MemWarp sets itself apart from existing approaches in two key ways.
First, it decouples feature learning from similarity matching by utilizing Laplacian pyramids to create residual deformation fields at each level of a Unet \cite{ronneberger2015u}, allowing it to capture deformations from coarse to fine. 
Second, unlike conventional learning-based methods that entangles features of moving and fixed images, MemWarp uses the fixed image’s feature map to steer the creation of dynamic filters. 
These filters, tailored to specific anatomical regions, improve the model's ability to manage discontinuities across different areas.
MemWarp's performance is validated on a public cardiac dataset \cite{bernard2018deep}, where it surpasses other state-of-the-art semi-supervised methods by a large margin.
The main findings of this study are as follows:
\begin{itemize}
\item Decoupling feature extraction from similarity matching yields registration accuracies on par with intertwined methods in unsupervised contexts;
\item This decoupling allows flexible use of fixed feature maps, leading to a memory network that retains dynamic filters specific to anatomical regions to promote local discontinuities;
\item MemWarp excels beyond all leading semi-supervised methods in registration accuracy, achieving a significant 7.1\% improvement in Dice score; it outperforms discontinuity-preserving models without needing the segmentation masks for inference that are typically required by these approaches.
\end{itemize}

%% file: docs/method.tex
\section{Methodology}

\subsection{Preliminaries}
Deformable image registration aims to establish voxel-level correspondences between a moving image $\mathbf{I}_m$ and a fixed image $\mathbf{I}_f$.
The spatial relationship is represented by $\mathbf{\phi}(x) = x + \mathbf{u}(x)$, where $x$ is a spatial location within the domain $\mathbf{\Omega} \subset \mathbf{R}^{H\times W\times D}$, and $\mathbf{u}(x)$ denotes the displacement vector at that location.
In unsupervised learning, a network $F_{\theta}$ is trained to predict the deformation field $\phi$, with its weights $\theta$ optimized by minimizing a composite loss function $\mathcal{L}$. 
This function combines metrics for dissimilarity between the warped moving image and the fixed image, and the smoothness of the deformation field: $\mathcal{L} = \mathcal{L}_{sim}(\mathbf{I}_f,\mathbf{I}_m \circ \phi) + \lambda \mathcal{L}_{reg}(\phi)$.
Here, $\lambda$ serves to balance the smoothness constraint on the deformation field, with methods like the discontinuous regularizer proposed by Ng et al.~\cite{ng2020unsupervised} falling under this strategy.
Semi-supervised methods, including our MemWarp, introduce an additional Dice loss $\mathcal{L}_{dsc}(\mathbf{J}_f,\mathbf{J}_m\circ \phi)$ to assess the dissimilarity between the warped moving mask and the fixed mask.
Weakly-supervised models need mask inputs for the network $F_{\theta}$. 
For instance, DDIR \cite{chen2021deep} requires both moving and fixed masks, while textSCF \cite{chen2024spatially} requires only the fixed mask.

\subsection{Laplacian Pyramid Warping Network}

To decouple feature extraction from similarity matching, we develop a Laplacian pyramid warping network (LapWarp) that leverages residual deformation fields across multiple scales, from coarse to fine. 
Contrary to previous method LapIRN \cite{mok2020large,mok2021large}, which applies image pyramids directly to raw images, LapWarp performs warping on feature maps and allows for interactions at all levels of the pyramid.
This ensures stable training within its pyramid framework without requiring the warm starts or multi-stage coarse-to-fine training strategies. 

\textbf{Network Architecture:}
LapWarp deviates from classic Unet by stacking moving and fixed images across the batch dimension and employing a unique decoder structure. 
In each decoder level, moving image features are first warped using the previous level's field. 
A standard decoder layer then extracts features from both images as a batch, which a flow generator uses at each pyramid level to create the residual deformation field by re-stacking features along channels.

\begin{figure}[!t]
	\centering
        \includegraphics[width=1.0\columnwidth]{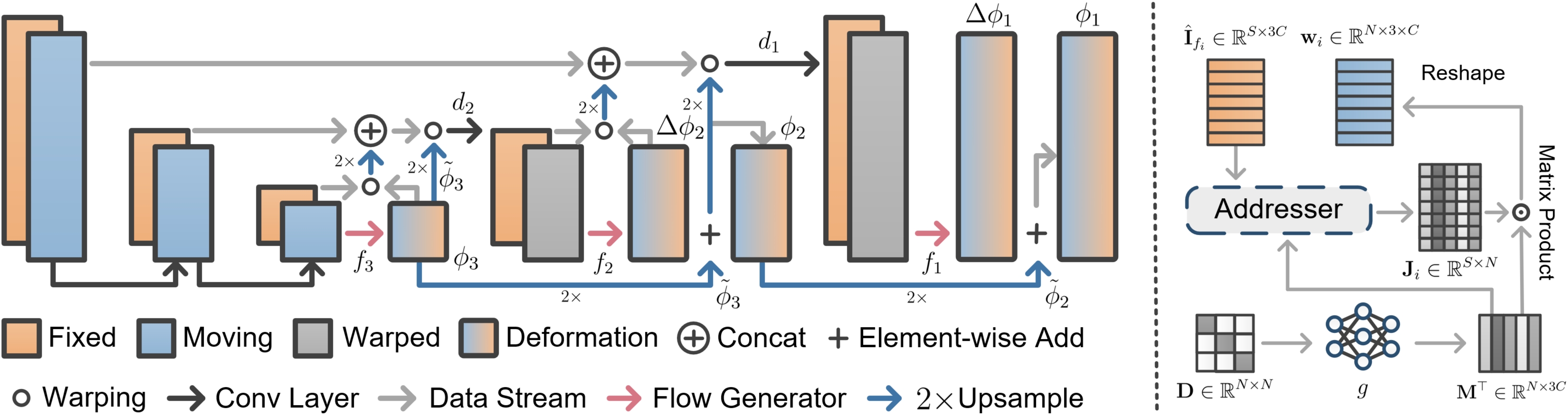}
        \caption{
        Schematic representation of the MemWarp framework. 
        The left panel depicts a 2-level LapWarp network employing Laplacian image pyramids; the right panel outlines the operation of the memory network.
        }
	\label{fig:lapwarp}
\end{figure}

Given \(n\) pyramid levels, we obtain \(n\) residual deformation fields, labeled from $\Delta \tilde{\phi}_{n}$ to $\Delta \tilde{\phi}_{1}$, and \(n+1\) total deformation fields, labeled from \(\phi_{n+1}\) to \(\phi_{1}\), with both sets following the convention that a larger index indicates a coarser level.
At level $i+1$, the feature maps $\hat{\mathbf{I}}_{m_{i+1}}$ and $\hat{\mathbf{I}}_{f_{i+1}}$ are generated by its decoder $d_{i+1}$.
These feature maps, stacked along the channel dimension, are processed by the flow generator $f_{i+1}$ to produce the residual deformation field $\Delta \phi_{i+1} = f_{i+1}(\hat{\mathbf{I}}_{m_{i+1}} \oplus \hat{\mathbf{I}}_{f_{i+1}})$. 
This residual field is then combined with the upsampled and scaled (by a factor of 2) deformation field $\tilde{\phi}_{i+2}$ from level $i+2$, resulting in the deformation field for level $i+1$, given by $\phi_{i+1} = \Delta \phi_{i+1} + \tilde{\phi}_{i+2}$. 
For the $i_{th}$ level, the encoder feature maps $\mathbf{I}_{m_i}$ and $\mathbf{I}_{f_i}$, together with upsampled decoder feature maps $\tilde{\mathbf{I}}_{m_{i+1}}$ and $\tilde{\mathbf{I}}_{f_{i+1}}$, undergo processing by the decoder layer $d_{i}$ and flow generator $f_{i}$ to produce this level's deformation field:  
\begin{align}
    \hat{\mathbf{I}}_{f_{i}} & = d_i(\mathbf{I}_{f_{i}} \oplus \tilde{\mathbf{I}}_{f_{i+1}}),  \nonumber \\
    \hat{\mathbf{I}}_{m_{i}} & = d_i((\mathbf{I}_{m_{i}} \circ \tilde{\phi}_{i+1}) \oplus (\tilde{\mathbf{I}}_{m_{i+1}} \circ \Delta \tilde{\phi}_{i+1})), \nonumber \\
    \Delta \phi_i & = f_i(\hat{\mathbf{I}}_{m_{i}} \oplus \hat{\mathbf{I}}_{f_{i}}), \nonumber \\
    \phi_i & = \Delta \phi_i + \tilde{\phi}_{i+1},
    \label{eq:lpn}
\end{align}
where $\Delta \tilde{\phi}_{i+1}$ and $\tilde{\phi}_{i+1}$ are the upsampled and scaled deformation fields from the previous level, $\Delta \phi_i$ denotes the residual deformation field at level $i$, and $\phi_i$ represents the cumulative deformation field at this level.
It's worth noting that when $i+1$ is the coarsest level, we treat $\tilde{\phi}_{i+2}$ as an identity grid (with zero displacement field).
Fig. \ref{fig:lapwarp} depicts a two-level LapWarp for visual illustration. 
Typically, the number of pyramid levels corresponds to the count of downsampling layers.
 
\subsection{Discontinuity-Preserving Deformable Registration}
DDIR \cite{chen2021deep} is the first neural network solution to generate high-quality, discontinuity-preserving deformation fields, but it requires segmentation masks for both training and inference, linking deformation field quality to segmentation accuracy. 
Additionally, DDIR's use of masks increases computational load by splitting image pairs per anatomical region. 
MemWarp tackles these challenges by incorporating a memory network \cite{sukhbaatar2015end} that adaptively learns prototypical feature representations for different anatomical regions. 
Empirical evidence suggests that learning such prototypical features is not feasible when features from moving and fixed images are entangled, which led to the development of LapWarp.

\textbf{Anatomical Filters:}
Typically, the flow generator uses convolutional or self/cross-attention layers as in transformers, ending with a single convolutional filter of kernel size 1 to produce the deformation field. 
Our approach replaces this filter with dynamic filters \cite{zhang2023spatially,chen2024spatially}, adapting to the voxel context based on fixed feature maps.
Given $x$ as a location vector within $\mathbf{\Omega} \subset \mathbb{R}^{H_i\times W_i \times D_i}$, let $\hat{\mathbf{I}}_{m_{i}}$ and $\hat{\mathbf{I}}_{f_{i}}$ represent the moving and fixed feature maps from the decoder at pyramid level $i$, respectively. 
The function $f_{i_c}$ denotes the operation of the convolutional layer. 
With $\mathbf{w}_i(x) \in \mathbb{R}^{C \times 3}$ as the designated filter at position $x$ in the flow generator, the residual displacement vector at $x$ is defined as $\Delta \mathbf{\phi}_i(x) = \mathbf{w}_i(x)^{\top}f_{i_c}(\hat{\mathbf{I}}_{m_{i}} \oplus \hat{\mathbf{I}}_{f_{i}})(x)$, where $f_{i_c}(\hat{\mathbf{I}}_{m_{i}} \oplus \hat{\mathbf{I}}_{f_{i}})(x) \in \mathbb{R}^{C}$ yields the context vector at $x$. 
Unlike the conventional approach that applies a uniform $\mathbf{w}_i(x)$ across all locations, our method allows for dynamic filter generation.

Filter generation involves a memory query, addressing, and reconstruction process.
The fixed feature vector $\hat{\mathbf{I}}_{f_{i}}(x)\in \mathbb{R}^{3C}$ acts as the query, with $\mathbf{M}\in \mathbb{R}^{3C\times N}$ representing the memory matrix containing $N$ slots, where $N$ denotes the number of anatomical regions.
Instead of storing $\mathbf{M}$ directly as learnable parameters, it is produced by a multi-layer perceptron (MLP).

\textbf{Memory Addressing \& Filter Generation:}
Define $\mathbf{D} \in \mathbb{R}^{N\times N}$ as a diagonal matrix filled with ones and $g$ as the MLP operation. 
The memory matrix $\mathbf{M}$ is derived as $\mathbf{M} \in \mathbb{R}^{3C\times N} = g(\mathbf{D})$. 
Utilizing the fixed feature map $\hat{\mathbf{I}}_{f_{i}}\in \mathbb{R}^{S_i\times 3C}$ ($S_i=H_i\times W_i \times D_i$) as the query, memory addressing and filter generation proceed as follows: 
\begin{align}
    \mathbf{J}_i & = \text{softmax}\left(\frac{\hat{\mathbf{I}}_{f_{i}}\mathbf{M}}{\|\mathbf{M}\|_{2, a_1}}\right), \label{eq:J_i} \\
    \mathbf{w}_i & = \text{reshape}(\mathbf{J}_i \mathbf{M}^{\top}),
\end{align}
where the division by $\| \mathbf{M} \|_{2, a_1}$ applies $L_2$ normalization along the $1_{st}$ axis of the tensor $\mathbf{M}$, and the softmax is then applied along the $2_{nd}$ axis of the tensor.
With $\mathbf{w}_i$ obtained, the deformation field is generated in accordance with Eq.~\eqref{eq:lpn}.
The reshape function transforms $\mathbf{w}_i \in \mathbb{R}^{S \times 3C}$ into $\mathbf{w}_i \in \mathbb{R}^{S \times 3 \times C}$.

\textbf{Anatomical Region Loss:}
\textbf{Anatomical Region Loss:}
The feature representation $\hat{\mathbf{I}}_{f_{i}}(x)$at pyramid level $i$ of the fixed image produces the memory-addressed $\mathbf{J}_i \in \mathbb{R}^{S\times N}$, which acts as a segmentation probabilities across $N$ regions.  
Across all pyramid levels, we apply Dice loss: $\mathcal{L}_{rgn} = \sum_i^{n} \text{DSC}(\text{up}(\mathbf{J}_i), \mathbf{J}_{f}) \times \frac{1}{2^{i-1}}$, where $\mathbf{J}_i$ is the network output, $\mathbf{J}_f$ is the fixed segmentation mask from the dataset, and the up function upsamples $\mathbf{J}_i$ to match $\mathbf{J}_f$'s resolution.

\subsection{Loss function \& Overall Framework}

The composite loss function for MemWarp is formulated as:
\begin{equation}
    \mathcal{L} = \mathcal{L}_{sim}(\mathbf{I}_f,\mathbf{I}_m \circ \phi) + \mathcal{L}_{dsc}(\mathbf{J}_f,\mathbf{J}_m \circ \phi) + \lambda_1 \mathcal{L}_{reg} + \mathcal{L}_{rgn},
    \label{eq:loss}
\end{equation}
with $\mathcal{L}_{reg}=\sum_{x\in\Omega}||{\nabla \mathbf{u}_i(x) }||^2$ ($\mathbf{u}_i(x)=\mathbf{\phi}_i(x)-x$) and $\lambda$ adjusting the smoothness regularization strength.
The framework of MemWarp aligns with traditional registration frameworks like VoxelMorph but introduces three critical adjustments: 1) Moving and fixed images are combined along the batch dimension; 2) Flow generators, enhanced by memory networks, supplement a conventional Unet, yielding a gradually warped moving image for each decoder level; 3) Deep supervision \cite{lee2015deeply} is employed on the memory-addressed tensors to encourage discontinuities across regions.

%% file: docs/experiment.tex
\section{Experiments \& Results}

We evaluate MemWarp's effectiveness using the ACDC dataset \cite{bernard2018deep}, which includes 150 subjects.
Each subject is provided with images from both End-diastole (ED) and End-systole (ES) phases alongside segmentation masks. 
For intra-subject registration, images from both ED to ES and ES to ED phases are required to be aligned, resulting in a total of 300 pairs ($[100+50] \times 2$). 
Of these, 170 pairs are allocated for training, 30 for validation, and the remaining 100 for testing. 
The distribution is stratified to ensure subjects with various diseases are evenly represented across training, validation, and testing phases, with no overlap of subjects between training or validation and testing.
All images undergo a min-max normalization to (0,1), are resampled to a voxel size of $1.8\times1.8\times10$ mm and adjusted to a size of $128\times128\times16$.

\subsection{Implementation Details \& Comparator Methods}

Experiments use Python 3.7 and PyTorch 1.9.0 \cite{paszke2019pytorch} on a machine equipped with an A100 GPU, and a 16-core CPU with 32GB RAM. 
Training employs the Adam optimizer with a learning rate of 4e-4, a batch size of 4, and cosine decay, running for 400 epochs. 
The Mean Square Error (MSE) serves as the similarity loss, complemented by L2 regularization on the spatial gradients of the deformation field ($\lambda=0.01$ in Eq.~\eqref{eq:loss}), following \cite{dalca2019unsupervised,balakrishnan2019voxelmorph}, with seven integration steps in the diffeomorphic layer.
For MemWarp, a diffeomorphic layer is used at all pyramid levels except the first. 
Other models apply only MSE loss, Dice Loss, and regularization as outlined in Eq.~\eqref{eq:loss}'s initial three terms.

\textbf{Comparator Methods:}
MemWarp is benchmarked against top learning-based models such as VoxelMorph \cite{balakrishnan2019voxelmorph}, TransMorph \cite{chen2022transmorph}, LKU-Net \cite{jia2022u}, and Slicer Network \cite{zhang2024slicer}, as well as DDIR \cite{chen2021deep} which is recognized for its discontinuity-preserving capabilities in cardiac registration. 
For DDIR, we employ the leading model nnFormer \cite{zhou2023nnformer} for segmentation, achieving a Dice score of 90.15\% on the test set.
Slicer Network is assessed with an added guidance loss per its original configuration, while MemWarp and the other models are tested under a consistent experimental framework.
We also include traditional methods like ANTs \cite{avants2011reproducible} and Demons \cite{vercauteren2007diffeomorphic}.
While MemWarp is model-agnostic, we utilize the backbone of LKU-Net in this implementation.

\textbf{Evaluation Metrics:}
Aligned with standard practices \cite{balakrishnan2019voxelmorph,chen2022transmorph}, our evaluation employs the Dice coefficient and the 95th percentile Hausdorff Distance (HD95) for anatomical alignment evaluation. 
HD95 values are averaged across all anatomical structures for individual subjects. Additionally, the standard deviation of the logarithm of the Jacobian determinant (SDlogJ) is utilized to evaluate the quality of diffeomorphism.

\begin{table}[t!]
\centering
\caption{Comparative analysis of MemWarp (LapWarp denotes a MemWarp variant wihtout the memory module) and other models on the test set of the ACDC dataset, with top performing metric in bold. 
Metrics include Average Dice (\%), RV Dice (\%), LVM Dice (\%), LVBP Dice (\%), HD95 (mm), and SDlogJ, with lower values preferred for HD95 and SDlogJ. 
For clarity, models are categorized as unsupervised (trained solely on raw images), semi-supervised (using segmentation masks in training), and weakly-supervised (requiring masks during both training and inference).
}

\resizebox{0.99\columnwidth}{!}{
\begin{tabular}{lccccccc}
\hline
\hline
Model   & Type & Avg. ($\%$)  & RV ($\%$) & LVM ($\%$) & LVBP ($\%$) & HD95 (mm) $\downarrow$ &SDlogJ $\downarrow$ \\ \hline
Initial & - & 58.14 & 64.50 & 48.33 & 61.60 & 11.95 & - \\
\hline
ANTs \cite{avants2011reproducible}  &\multirow{3}{*}{\shortstack[l]{~Traditional}} & 71.04 & 68.61 & 67.53 & 76.96 & 13.15 & 0.056\\
Demons \cite{vercauteren2007diffeomorphic} & & 72.37 & 70.85 & 69.34 & 76.93 & 11.46 &  0.031\\
Bspline \cite{marstal2016simpleelastix} & & 74.36 & 72.18 & 71.68 & 79.22 & 11.18 & 0.030\\
\hline
TransMorph \cite{chen2022transmorph} & \multirow{4}{*}{\shortstack[l]{~Unsupervised}}& 74.97 & 73.08   & 71.49    & 80.34  & 9.44 & 0.045 \\
VoxelMorph \cite{balakrishnan2019voxelmorph}	&  & 75.26 & 73.10   & 71.80  & 80.88  & 9.33 & 0.044 \\
LKU-Net \cite{chen2022transmorph}  && 76.53 & 74.25   & 73.23    & 82.12  & 9.13 & 0.049  \\
Slicer Network \cite{zhang2024slicer}  && 79.52 & 77.83   & 76.80    & 83.93  & 8.21 & 0.044  \\
\textbf{LapWarp (ours)} && 77.25 & 75.86 & 73.92 & 81.99 & 9.23	& 0.074 \\
\hline
TransMorph \cite{chen2022transmorph} &\multirow{4}{*}{\shortstack[l]{~Semi-supervised}}& 81.08 & 81.73   & 75.27    & 86.23  & 7.51 & 0.091 \\
VoxelMorph \cite{balakrishnan2019voxelmorph}	&  & 81.34 & 82.03   & 75.35   & 86.64  & 6.87  & 0.082  \\
LKU-Net \cite{chen2022transmorph}  && 83.08 & 84.59   & 77.24    & 87.39  &  6.43 & 0.099 \\
Slicer Network \cite{zhang2024slicer}  && 83.68 & 84.94   & 77.97    & 88.12  & 6.10 & 0.099  \\
\textbf{MemWarp (ours)} && \textbf{89.61}	& 89.30 & \textbf{86.49}	& \textbf{93.04}	&  \textbf{3.93}	& 0.149 \\
\hline
DDIR \cite{chen2021deep} & \multirow{1}{*}{\shortstack[l]{~Weakly-supervised}}  & 88.03 & \textbf{90.02 }& 85.47 & 87.61 & 9.91 & 0.121 \\
\hline
\hline
\end{tabular}
}
\label{tab:acdc}
\end{table}

\subsection{Results \& Analysis}

\textbf{Registration Accuracy:}
Table \ref{tab:acdc} illustrates that all methods produce smooth displacement fields with low SDlogJ values; however, increased SDlogJ alongside higher Dice scores indicates inherent discontinuities in cardiac alignments. 
Among unsupervised learning-based models, all outperform traditional methods, with Slicer Network at the forefront due to its large effective receptive field (ERF) and TransMorph lagging, likely hindered by insufficient training data for its transformer architecture. 
In semi-/weakly-supervised contexts, MemWarp and DDIR, which focus on preserving discontinuities, lead the pack. 
Despite Slicer Network's strong unsupervised performance, its limited handling of local discontinuities relegates it behind MemWarp. 
Notably, MemWarp surpasses all semi-supervised methods with a significant 7.1\% Dice score gain.
DDIR, while competitive, shows potential drawbacks from segmentation inaccuracies, indicated by a higher HD95 value.

\textbf{Ablation Analysis:}
Table \ref{tab:ablation} details our ablation study, examining the impact of the Laplacian pyramid, the memory network, and the inclusion of Dice loss.
The base model, labeled as $\# 1$, functions as the backbone network in the unsupervised setting, with enhancements observed in $\# 2$ upon integrating the Laplacian pyramid. 
In the semi-supervised scenario, the memory network generates segmentation masks comparable in accuracy to top-tier models like nnFormer \cite{zhou2023nnformer}, utilized in DDIR's mask generation (89.68\% vs 90.15\%). 
Yet, we observe that excluding $\mathcal{L}_{dsc}$ can tilt the network's focus towards segmentation, which consequently degrades registration accuracy and the smoothness of the displacement field, as evidenced by $\# 4$.
Moreover, comparing $\# 4$ and $\# 5$, registration accuracy improves even in the absence of the Laplacian pyramid when Dice loss is included. 
The optimal performance in both registration and segmentation is achieved when all three components are included, as with $\# 6$.

% \begin{figure}[!t]
%     \centering
%     \subfloat[TransMorph]{\includegraphics[width=.22\columnwidth]{figs/TransMorph_erf.pdf} \label{fig:TransMorph_erf}}
%     \subfloat[VoxelMorph]{\includegraphics[width=.22\columnwidth]{figs/VoxelMorph.pdf} \label{fig:VoxelMorph}}
%     \subfloat[LKU-Net]{\includegraphics[width=.22\columnwidth]
%     {figs/LKU-Net.pdf} \label{fig:LKU-Net}}
%     \subfloat[MemWarp]{\includegraphics[width=.22\columnwidth]
%     {figs/MemWarp.pdf} \label{fig:MemWarp}}
%     \caption{ 
%         ERF visualizations \protect\cite{luo2016understanding}.
%         Darker and wider regions indicate a larger ERF. 
%     }
%     \label{fig:erf_examples}
% \end{figure}

\begin{figure}[!t]
	\centering
        \includegraphics[width=1.0\columnwidth]{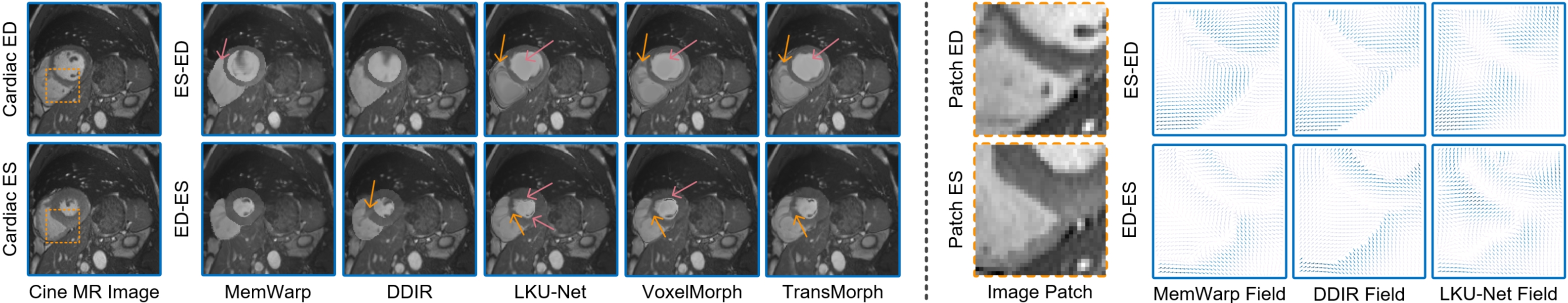}
        \caption{
        Comparative visualization of MemWarp against other methods on cardiac MR images, highlighting deformable registration across ED $\Leftrightarrow$ ES phases. 
        Pink arrows show omitted trabeculations; orange arrows identify artifacts. 
        The right panel focuses on deformation fields outlined by the left panel's yellow dash, with arrow darkness indicating displacement magnitude.
        }
	\label{fig:image_examples}
\end{figure}
% \textbf{Effective Receptive Field (ERF) Analysis:}
% Fig. \ref{fig:erf_examples} displays heatmaps of the ERF \cite{luo2016understanding} for different models. 
% The visual patterns suggest that a denser and broader ERF benefits the registration task, as observed with MemWarp, which exhibits the most intense and expansive ERF coverage. 
% Additionally, a smaller, more intensely heated area is preferable to a larger, less intense one, as the comparison between VoxelMorph and LKU-Net illustrates.

\textbf{Qualitative Analysis:}
Fig. \ref{fig:image_examples} showcases MemWarp's qualitative performance. 
Notably, MemWarp minimizes artifacts and consistently captures cardiac structures like trabeculations. 
DDIR's artifacts, particularly between the ventricles, may stem from segmentation inaccuracies. 
MemWarp and DDIR both display clear organ boundary discontinuities, in contrast to LKU-Net's blending of these regions.
MemWarp also manages background deformations adeptly, avoiding DDIR's tendency to reduce displacement magnitude. 
Within organs, MemWarp finely tunes deformation with respect to the underlying texture instead of overly smoothing the field.

\subsection{Discussions}

Let \(\mathbf{I}_{m_i}\) and \(\mathbf{I}_{f_i}\) be the feature maps of moving and fixed images at pyramid level \(i\). 
MemWarp operates under the assumption that the 'brightness' at any given location $p \in \Omega$ in \(\mathbf{I}_{f_i}\) remains constant compared to moving image \cite{horn1981determining}, which is formulated as:
\begin{equation}
    \nabla \mathbf{I}_{f_i}(p) \cdot \mathbf{u}(p) = \mathbf{I}_{m_i}(p) - \mathbf{I}_{f_i}(p),
    \label{eq:horn_schunck}
\end{equation}
where \(\nabla \mathbf{I}_{f_i}(p) = \left[ \frac{\partial \mathbf{I}_{f_i}}{\partial x}(p), \frac{\partial \mathbf{I}_{f_i}}{\partial y}(p), \frac{\partial \mathbf{I}_{f_i}}{\partial z}(p) \right]^T\). 
Eq.~\eqref{eq:horn_schunck} holds provided that the magnitude of \(\mathbf{u}(x)\) is less than one voxel.
In the MemWarp framework, we employ an \(n\)-level Laplacian image pyramid to ensure \(2^{(n-1)} > d_{max}\), where \(d_{max}\) is the maximum possible displacement magnitude.
This setup ensures that the coarsest level meets the conditions of Eq.~\eqref{eq:horn_schunck}, with each finer level processing a pre-warped moving image, thus maintaining the model's assumption throughout all levels.

Based on the assumption, we've implemented two major modifications in neural network architecture to enhance registration performance. 
First, we decouple feature learning from flow estimation. 
Unlike traditional registration networks that combine moving and fixed images at the network's input, MemWarp employs a U-net for feature extraction and adds a simple convolution layer at each pyramid level to compute flow and performs warping, ensuring each level satisfies Eq.~\eqref{eq:horn_schunck}. 
Second, the smoothness requirement of Eq.~\eqref{eq:horn_schunck} aligns well with features derived from segmentation networks, as segmentation can be regarded as the ultimate form of image harmonization \cite{blake1987visual}.
This reinforces that effective segmentation features are equally beneficial for registration. 
Consequently, MemWarp uses fixed feature maps to steer dynamic filter creation, enhancing feature map smoothness within organs and preserving local discontinuities across boundaries.

\begin{table}[t]
\centering
\caption{Ablation results outlining the individual and combined contributions of the Laplacian pyramid, memory network, and Dice loss to the performance of our model, achieving optimal outcomes when all three modules are employed.}
\label{tab:ablation}
\resizebox{0.99\columnwidth}{!}{
\begin{tabular}{ccccccccc}
\hline
\hline
Model ID & ~Pyramid~ & ~Dice Loss~ & ~Memory~ & Type & Avg. (\%) & HD95 (mm) $\downarrow$ & SDlogJ $\downarrow$ & Seg Dice (\%) \\
\hline
\# 1 & $\times$ & $\times$ & $\times$ & \multirow{2}{*}{\shortstack[l]{~Unsupervised}} & 76.53 & 9.13 & 0.049 & - \\
\# 2 & $\checkmark$ & $\times$ & $\times$ & & 77.25 & 9.23 & 0.074 & - \\
\hline
\# 3 & $\times$ & $\checkmark$ & $\times$ & \multirow{4}{*}{\shortstack[l]{~Semi-supervised}} & 83.08 & 6.43 & 0.099 & - \\
\# 4 & $\checkmark$ & $\times$ & $\checkmark$ & & 74.81 & 9.26 & 0.950 & 89.06 \\
\# 5 & $\times$  & $\checkmark$& $\checkmark$ & & 85.87 & 5.32 & 0.085 & 87.76 \\
\# 6 & $\checkmark$ & $\checkmark$ & $\checkmark$ & & 89.61 & 3.93 & 0.149 & 89.68 \\
\hline
\hline
\end{tabular}
}
\end{table}

%% file: docs/conclusion.tex
\section{Conclusions}

In conclusion, MemWarp establishes a new benchmark for cardiac registration, outperforming existing methods by effectively preserving essential anatomical details and reducing artifacts. 
Its success hinges on two pivotal elements: the decoupling of moving and fixed feature maps via LapWarp, and the memory network's use of region loss for maintaining discontinuities across boundaries. 
MemWarp's effectiveness is validated by a significant 7.1\% Dice score enhancement over the nearest semi-supervised competitors. 
Moreover, MemWarp uniquely addresses discontinuities without needing segmentation masks at inference, yet it can still generate segmentation masks comparable to top segmentation methods. 